\documentstyle[12pt]{article}

\newcommand{\beq}{\begin{equation}}
\newcommand{\eeq}{\end{equation}}
\newcommand{\id}{i\!\!\not\!\partial}
\newcommand{\as}{\not\!\! A}
\newcommand{\ps}{\not\! p}
\newcommand{\ks}{\not\! k}
\newcommand{\D}{{\cal{D}}}
\newcommand{\dv}{d^3x}
\newcommand{\Z}{{\cal Z}}
\newcommand{\N}{{\cal N}}

\begin{document}
\title
{Parity Violation in the Three Dimensional Thirring Model}
\author{
G.Rossini\thanks{CONICET, Argentina} ~and F.A.Schaposnik\thanks{Investigador
CICBA,
Argentina}\\ {\normalsize\it Departamento de F\'\i sica, Universidad
Nacional de La Plata}\\ {\normalsize\it C.C. 67, (1900) La Plata,
Argentina}}
\date{}

\maketitle


\begin{abstract}
\normalsize We discuss parity violation in the 3-dimensional (N flavour)
Thirring model. We find that the ground state fermion current in
a background gauge field does not posses a well defined parity transformation.
We also investigate the connection between parity violation and
fermion mass generation, proving that radiative corrections
force the fermions to be massive.
\end{abstract}
\newpage
\pagenumbering{arabic}
Planar theories in three dimensional space-time \cite{Jac1} show a variety
of interesting phenomena relevant not only for Quantum Field Theory but
also for Condensed Matter physics.

A first important feature of three dimensional kinematics concern the
possibility of giving a (topological) mass to the vector field by including
an unconventional term in gauge field Lagrangians \cite{Jac2}-\cite{Scho}.
This term, of topological origin, is the Chern-Simons (CS) secondary
characteristic.

Now, a salient property of the CS term is that it violates P and T
invariance. Since the same happens with the mass term for a (two-component)
Dirac spinor in three dimensions, it is natural to expect interesting
connections between both masses in three dimensional gauge theories coupled
to fermions. Indeed, in refs.\cite{Jac2}-\cite{Jac3} it was shown that if
any of the two mass terms is inserted in the Lagrangian, the other is then
induced by radiative corrections.

Later on, it was also shown that even massless fermions, when coupled to
gauge fields, generate a CS term \cite{NS}-\cite{Red}. Originally, this
effect was  thought as a consequence of the introduction of a fermion mass
term whithin the
(Pauli-Villars) regularization procedure \cite{Red}. However, the
occurrence of the CS term was confirmed in
ref.\cite{GMSS} using
the $\zeta$-function approach, where no regulating mass-term is added at
any stage of the calculations. In fact, the violation of parity in odd
dimensions  (and, consequently, the { generation}  of a CS term) is
analogous to the non-conservation of the axial current in even dimensions:
the imposition of gauge invariance produces in both cases an anomaly for a
symmetry (parity, chiral symmetry) of the original action for {\it any}
sensible regularization.

As remarked  above, the gauge field and fermion mass terms belong together
since both violate P and T; in perturbation theory, one can be generated
from the other. Indeed, by including a bare CS term in the
$QED_3$ Lagrangian,  Deser, Jackiw and Templeton \cite{Jac3}  showed that
even if the bare fermion mass
is set to zero, the physical (renormalized) fermion mass still does not
vanish
(see also \cite{Y-C}).

In connection with these interesting phenomena, gauge models containing CS
terms have attracted special attention in different areas. The possibility
of giving a concrete realization of exotic spin and statistics popularized
CS models in
condensed matter problems like the Hall effect and high-Tc
superconductivity
\cite{Frad}. Connections between pure  CS theory with rational conformal
field theories in two dimensions as well as with knot theory were also
discovered \cite{Wi}
and considerable developements have been achieved in this area \cite{BBR}.

We discuss in this letter a three dimensional purely fermionic model,
the ($N$ flavour) Thirring model which, as we shall see, shares many of the
attractive features described above. In fact, by introducing a vector
auxiliary field to eliminate the quartic interaction, the appearence of a
CS term is made apparent together with the breaking of parity and the
consequent fermion mass generation.

Four fermion interaction models in 3 dimensions with $N$ flavours are
known to be renormalizable in the $1/N$ expansion \cite{Ros}. Precisely,
renormalizability and dynamical mass generation have been investigated in
the  Thirring model within the $1/N$ approximation, by
introducing scalar and vector fields to split the current-current
interaction \cite{Kras}-\cite{ACh}.
Experience with  2-dimensional Thirring and Chiral Gross-Neveu
models  has shown the convenience of using vector fields (instead of scalar
fields, as originally done in \cite{GN}) to achieve this splitting when
exploring properties where symmetries play a crucial r\^ole
\cite{FGS}-\cite{MS}. In this way, the decoupling of massless excitation as
well as the 2-dimensional ``almost long-range order'', predicted by means of
the $1/N$ expansion \cite{Wi2} can be clearly seen without any kind of
approximation.   Following this route we shall show that parity violation,
through the appearence of a CS terms for the auxiliary field,
takes place in the 3-dimensional Thirring model. From this, generation of a
fermion mass can be inferred and in fact we shall explicitly see that
parity violation requires the fermion to be massive.

We start from the 3-dimensional (Euclidean) Thirring model Lagrangian:

\beq
{\cal L}_{Th}= \bar\psi^i \id \psi^i -\frac{g^2}{2N}J^{\mu}J_{\mu}
\label{1}
\eeq
where $\psi^i$ are $N$ two-component Dirac spinors and $J^{\mu}$ the $U(1)$
current,
\beq
J^{\mu}=\bar\psi^i\gamma^{\mu}\psi^i.
\label{2}
\eeq
The coupling constant $g^2$ has dimensions of inverse mass.

The partition function for the theory is defined as
\beq
\Z[B_{\mu}] = \N_1 \int \D\bar\psi \D\psi \exp[
-\int(\bar\psi^i \id \psi^i -\frac{g^2}{2N}J^{\mu}J_{\mu}+J^{\mu}B_{\mu})
\dv]
\label{3}
\eeq
where $B_{\mu}$ is an external source and $\N_1$ is a normalization
constant.

In order to evaluate $\Z[B_{\mu}] $ we first eliminate the quartic
interaction by introducing a vector field $A_{\mu}$ through the identity
\beq
\exp(\int \frac{g^2}{2N}J^{\mu}J_{\mu} \dv)=\N_2 \int\D A_{\mu}
\exp[-\int (\frac{1}{2}A^{\mu}A_{\mu}+\frac{g}{\sqrt{N}}J^{\mu}A_{\mu})\dv]
\label{4}
\eeq
(here ${\cal{N}}_2$ is some normalization constant) so that the partition
function
becomes
\beq
\Z[B_{\mu}] = \N \int \D\bar\psi \D\psi \D A_{\mu} \exp[
-\int(\bar\psi^i (\id + \not\!\! B + \frac{g}{\sqrt{N}}\as)\psi^i
+\frac{1}{2} A^{\mu}A_{\mu})\dv].
\label{5}
\eeq
Before proceeding to integrate out the fermions we shift the vector field
in the form
\beq
\frac{g}{\sqrt{N}}A_{\mu}+B_{\mu} \to \frac{g}{\sqrt{N}}A_{\mu}
\label{6}
\eeq
so that we have
\begin{eqnarray}
\Z[B_{\mu}] &=& \N \int \D\bar\psi \D\psi \D A_{\mu} \exp[
-\int\bar\psi^i (\id + \frac{g}{\sqrt{N}}\as)\psi^i \dv] \times \nonumber
\\
 & &\exp [-\frac{1}{2}
\int (A_{\mu}-\frac{\sqrt{N}}{g}B_{\mu} )^2 \dv].
\label{7}
\end{eqnarray}
The fermionic path-integral gives, as usual, a determinant
\beq
\int \D\bar\psi \D\psi \exp(-\int\bar\psi^i (\id +
\frac{g}{\sqrt{N}}\as)\psi^i \dv) = det^N (\id+\frac{g}{\sqrt{N}}\as)
\label{8}
\eeq
which, as it is by now well known, has a parity violating contribution in
the form of a CS term. Indeed, it has been proven
\cite{NS}-\cite{GMSS} that the fermionic determinant takes the form
\beq
\log det (\id+\frac{g}{\sqrt{N}}\as) = \pm \frac{ig^2}{16\pi N}
\int\epsilon_{\mu\nu\alpha} F^{\mu\nu} A^{\alpha} \dv + I_{PC}[A_{\mu}]
\label{9}
\eeq
where $I_{PC}$ stands for parity conserving terms.

Some comments are here in order:
\begin{enumerate}

\item The result (\ref{9}) for massless fermions, originally obtained using
Pauli-Villars regularization \cite{Red},
has been reproduced by a variety of regularizations. In particular, in
ref.\ \cite{GMSS} we have shown the appearence of the CS term by
using the $\zeta$-function, a regularization scheme which makes unnecessary
the introduction of a regulating mass for fermions. This approach then
shows that the emergence of a (parity violating)  CS term is not just a
byproduct of a regularization scheme which (as the Pauli-Villars one),
breaks explicitly parity (through the introduction of regulating fermion
mass terms) but signals the impossibility of mantaining in three
dimensional gauge theories both gauge and parity invariance.

\item  More important for our purpose is to discuss the origin of double
sign in (\ref{9}). Within the Pauli-Villars scheme
it arises due to an ambiguity when taking the limit of the regulator
fermion mass going to zero. Concerning the $\zeta$-function approach,
the sign ambiguity can be traced back to the choice of an integration path
$\Gamma$ in the complex plane, necessary for defining the complex powers of
the Dirac operator \cite{Mat}. In odd dimensions, the choice
of $\Gamma$ in the upper (lower) half plane results in a positive
(negative) overall sign (see ref.\ \cite{GMSS} for details).

Again, the sign ambiguity is not a byproduct of a particular regularization
scheme but an intrinsic feature of regularization in odd
dimensional spaces.

We then consider that the only consistent way of
taking into account the sign ambiguity of the CS term arising from each one
of the N fermion determinants is to assign {\it the same sign} to all of
them. Any other alternative, as for example to take (for even $N$) one
half of ``$+$'' signs and one half of ``$-$'' signs (so that the overall CS
term would be absent, as advocated in \cite{HP}) would mean that
one has to define complex powers of
the Dirac operator in different ways for different fermion species, this
being in our opinion not mathematically sound.

\item The result (\ref{9}) was obtained in \cite{Red} by computing the
fermion current at one (fermion) loop; it is argued that  no further
radiative corrections arise from higher order loops. Calculations were
performed in that work for a constant and static $F_{\mu \nu}$. In
\cite{GMSS} the CS term was obtained in a
{\it non-perturbative}  way and for {\it arbitrary} $F_{\mu \nu}$.

\end{enumerate}

We shall not consider in what follows the (parity conserving) terms
included in
$I_{PC}[A_{\mu}]$ since they will play no r\^{o}le in the analysis of
parity
violation. Up to one loop one  has for example \cite{Red},
\beq
I_{PC}[A_{\mu}]= \frac{\zeta(3/2)}{\pi^2} \int
(\frac{g}{2\sqrt{N}}{F}_{\mu\nu}^2)^{3/2}\dv .
\label{10}
\eeq

We use now the result (\ref{9}) to write the partition function (\ref{7})
in the form
\beq
\Z[B_{\mu}] = \N \exp(-\frac{1}{2}\frac{N}{g^2}\int B_{\mu}B^{\mu} \dv)
\int \D A_{\mu} \exp(-S_{eff}[A_{\mu}])
\label{12}
\eeq
where $S_{eff}[A_{\mu}]$ is given by
\beq
S_{eff}=\frac{1}{2}\int A_{\mu}S^{\mu\nu}A_{\nu}\dv
-\frac{\sqrt{N}}{g}\int A_{\mu}B^{\mu}\dv
\label{13}
\eeq
and
\beq
S^{\mu\nu}=\delta^{\mu\nu} \mp
\frac{ig^2}{4\pi}\epsilon^{\mu\alpha\nu}\partial_{\alpha}.
\label{14}
\eeq
Performing the gaussian integral one readily obtains
\beq
\Z[B_{\mu}] = \N \exp(-\frac{1}{2}\frac{N}{g^2}\int B_{\mu}B^{\mu} \dv)
\exp(\frac{1}{2}\frac{N}{g^2}\int B_{\mu}G^{\mu\nu}B_{\nu} \dv d^3y)
\label{15}
\eeq
where $G^{\mu\nu}$ is the Green function for $S_{\mu\nu}$, whose Fourier
transform is
\beq
{\bar G}^{\nu\alpha}(k)=\frac{1}{1+a^2 k^2}(\delta^{\nu\alpha} +
{a^2}k^{\nu}k^{\alpha} \mp
{a}\epsilon^{\nu\rho\alpha}
k_{\rho}). \label{19a}
\eeq
Here $a = {g^2}/{4\pi}$.
In coordinate space  $G^{\nu\alpha}$ reads
\beq
G^{\nu\alpha}(x)=\frac{4\pi^3}{g^4}\frac{\exp(-\frac{|x|}{a})}{|x|}
\delta^{\nu\alpha}
-\frac{1}{4\pi}
\partial^{\nu}\partial^{\alpha}(\frac{\exp(-\frac{|x|}{a})}{|x|})
\pm
\frac{i\pi}{g^2}\epsilon^{\nu\rho\alpha}\partial_{\rho}
(\frac{\exp(-\frac{|x|}{a})}{|x|}).
\label{19}
\eeq

Let us first compute the fermion current expectation value in the
$B_{\mu}$ background:
\beq
\langle J_{\mu}(x)\rangle_B =
-\frac{1}{\Z}\frac{\delta \Z}{\delta B^{\mu}}=
\frac{N}{g^2}B_{\mu}(x) -\frac{N}{g^2}\int G_{\mu\alpha}(x-y)B^{\alpha}(y)
d^3y.
\label{20}
\eeq
The parity violating term in $J_{\mu}$ takes then the form
\beq
\langle J_{\mu}^{PV}\rangle = \pm \frac{ N}{g^4}
\int d^3y\frac{\exp(-{|x-y|}/{a})}{|x-y|}{^{*}{}F_{\mu}(y)} \label{21}
\eeq
where
\beq
^*F_{\mu}(x)=i \epsilon_{\mu\rho\alpha} \partial^{\rho}B^{\alpha}(x).
\label{21a}
\eeq
We can readily check that in the $g^2 \to 0$ limit we recover the well
known
result for non self-interacting fermions \cite{NS}-\cite{GMSS}. To this end
we note that the (three dimensional)
Dirac $\delta$-function can be represented as
\beq
\lim_{a\to 0} \frac{1}{4\pi a^2}\frac{\exp(-{|x-y|}/{|a|})}{|x-y|}=
\delta^{(3)}(x-y),
\eeq
so that one gets in the $g^2 \to 0$ limit:
\beq
\langle J_{\mu}^{PV}\rangle_{g^2\to 0} = \pm \frac{N}{4\pi}{^{*}F_{\mu}}
\eeq
in complete agreement with the non self-interacting result
\cite{NS}-\cite{GMSS}.

The Thirring model Lagrangian  ${\cal{L}}_{Th}$ defined in eq.(\ref{1})
is invariant under parity. We have found however that at the quantum level
a physical quantity, the ground state current in a gauge-field background
$\langle J_{\mu}\rangle_B$,  does not posses a well defined parity
transformation since there is a contribution $\langle
J_{\mu}^{PV}\rangle_B$, given by (\ref{21}), which transforms as a
pseudovector. Hence, parity is spontaneously broken. One should note that
although anomalous, $\langle J_{\mu}^{PV}\rangle$ is conserved, as
one easily checks from eqs.(\ref{21})-(\ref{21a}).

Our result (\ref{21}) is nothing but the extension to the case where there
is a fermion
self-interaction of those in refs.\
\cite{Jac2}-\cite{Red}, where parity violation through the generation of a
CS term
was discovered
for $QED_3$ and $QCD_3$.
Let us insist that we have adopted the point of view that the $N$
fermion contributions have to be considered with the same sign at the
light of the $\zeta$-function regularization analysis.

\vspace{0.2 cm}
Let us now show that the resulting
parity violation renders the fermion massive. To this end, instead of the
partition function (\ref{5}) we consider
\beq
\Z[\bar \eta, \eta] =\N\!\!\int \D\bar\psi \D\psi \D A_{\mu} \exp[
-\!\int(\bar\psi^i (\id + \frac{g}{\sqrt{N}}\as)\psi^i  +\bar\eta \psi +
 \bar\psi \eta
+\frac{1}{2} A^{\mu}A_{\mu})\dv]
\label{5B}
\eeq
where $\bar \eta$ and $\eta$  are anticommuting sources.   As before, we
integrate out the fermions and solely consider the parity violating
contribution to the fermion determinant. We thus have
\begin{eqnarray}
\Z[\bar \eta, \eta]  = \int \D A_{\mu}& &  \exp[\int (-\frac{1}{2}
A_{\mu} A^{\mu}
 \pm \frac{ig^2}{16\pi^2} \int\epsilon_{\mu\nu\alpha} F^{\mu\nu}
A^{\alpha} )\dv ]\times \nonumber \\
& & \exp[\int\bar \eta(x) G_F(x-y)\eta(y) d^3xd^3y]\label{11B}
\end{eqnarray}
Here,  $G_F(x-y)$ is the Dirac operator Green function,  \beq G_F(x-y) =
\frac{1}{\id + \frac{g}{\sqrt{N}}\as} \delta^3(x-y). \label{12B} \eeq  We
are now ready to evaluate the fermion propagator,
\beq {S_F(x-y) =
\langle\psi(x)\bar\psi(y)\rangle =
\frac{1}{\Z}\frac{\delta^2\Z}{\delta\bar\eta(x)\eta(y)}}
\left| \rule[-0.4cm]{0cm}{.4cm}   _{\bar \eta =\eta= 0}\right., \label{13B}
\eeq
which can be written in the form
\beq S_F(x-y) = \int \D A_{\mu}G_F(x-y)\exp[ \int (-\frac{1}{2}
A_{\mu} A^{\mu}
\pm \frac{ig^2}{16\pi} \epsilon_{\mu\nu\alpha} F^{\mu\nu}
A^{\alpha} )\dv ]  \label{14B}
\eeq
We now expand $G_F$ in powers of  $g/\sqrt N$, integrate
over $A_{\mu}$ and Fourier transform, getting
\beq
S_F(p)=\frac{-1}{\ps}+\frac{1}{N}\frac{-1}{\ps}\Sigma(p)\frac{-1}{\ps}
+O(1/N^2)
\label{29}
\eeq
with the fermion self-energy given by
\beq
\Sigma(p)=g^2\int\frac{d^3k}{(2\pi)^3}\gamma_{\mu}\frac{1}{\ks -\ps}
\gamma_{\nu}{\bar G}^{\mu\nu}(k).
\label{30}
\eeq
Note that the $A_{\mu}$ propagator ${\bar G}^{\mu\nu}(k)$
(given by eq.(\ref{19a})) depends on
$g^2$ and that our approximation corresponds effectively to a
$1/N$ expansion.

Now, from renormalization theory, one knows that the fermion physical mass
is determined by the equation \cite{NJL}
\beq
m={-\frac{1}{N}\Sigma(p)}\vert_{\not p = m}.
\label{31}
\eeq
To order $1/N$, we can set $m=0$ in the r.h.s.\ of eq.(\ref{31}) so that we get
\beq
m={-\frac{1}{N}\Sigma(p)}\vert_{\not p = 0}= \mp\frac{1}{N\pi^2}
\int d^3k\frac{1}{k^2+(\frac{4\pi}{g^2})^2}
\label{32}
\eeq
a result which picks as sole contribution that coming from the parity
violating term in $\bar G_{\mu\nu}$.
As expected, there is a linear divergence which will be handled by introducing
a
Pauli-Villars massive vector field (with mass $M$). Then,
the regularized form of  eq.(\ref{32}) takes the form
\beq
m=\mp\frac{2}{N}|M|+O(1/M).
\label{B}
\eeq

It has been established, within the $1/N$
approximation \cite{Ros}-\cite{Gom}, that the 3-dimensional Thirring
model is renormalizable. We can then make sense from
the divergent result (\ref{B}) adopting the following
viewpoint \cite{NJL}: as parity is violated there is no reason
for the fermion
to remain massless since it eventually acquires a (divergent) mass
(given by eq.(\ref{B})). One should then start from a Lagrangian with a
(parity violating) fermion bare mass term and see whether radiative
corrections yield a finite renormalized mass.

Now, if one starts with fermions having a bare mass $m_0$, then
eq.(\ref{9}) should be replaced by \cite{NS,Red}
\beq
\log det(\id+m_0+\frac{g}{\sqrt{N}}\as) = \frac{m_0}{|m_0|}
\frac{ig^2}{16\pi N}
\int\epsilon_{\mu\nu\alpha} F^{\mu\nu} A^{\alpha} \dv + I_{PC}[A_{\mu},m_0]
\label{33}
\eeq
Notice that the double sign arising in the massless case has been traded
by the sign of the bare mass.

One can now repeat all the steps leading
to eq.(\ref{29}). For the
renormalized mass $m$, instead of eq.(\ref{31}), one has \cite{NJL}
\beq
m=m_0-{\frac{1}{N}\Sigma'(p)}\vert_{\not p = m}
\label{34}
\eeq
where $\Sigma'(p)$ is now the self-energy corresponding to massive fermions.
Taking for simplicity $m_0$ to be of order $1/N$, one can simply write
\beq
m=m_0-{\frac{1}{N}\Sigma(p)}\vert_{\not p = 0}.
\label{35}
\eeq
{}From this we get, after Pauli-Villar regularization of $\Sigma(p)$,
\beq
m=m_0 -\frac{m_0}{|m_0|}\frac{2}{N}|M|+O(1/M).
\label{36}
\eeq
This result is highly non trivial in the sense that the opposite
signs of the two terms in (\ref{36}) make it possible to
define a finite value for the fermion mass. Had both contributions had
the same sign, one would have never attained a finite value
for the fermion mass through radiative corrections.
We consider this result as a proof of the fact that, due to parity
violation,
fermions do acquire a mass in the 3-dimensional Thirring model.


\vspace{0.2cm}
In summary, we have analysed parity violation in the 3-dimensional
(N flavour) Thirring model. We have found that the
ground state current in a gauge field background does not posses
a well defined parity transformation.  We have also studied the connection
between parity violation and mass generation for fermions.
By evaluating the fermion self-energy we have shown that
radiative corrections force the fermion to be massive.

Our approach is based on the introduction of an auxiliary vector
field followed by integration of fermions. This makes apparent
in a very simple way parity violation through the occurrence of
a CS term. It is important to notice that the route we
followed takes into account parity violating contributions
{\em exactly}. Indeed, the $\zeta$-function approach to the evaluation
of the fermionic determinant is not just a one loop approximation
but a complete result for parity violating terms \cite{GMSS} (of course the
parity conserving terms cannot be evaluated in a closed form).

The results above should be compared with those of ref.
\cite{Gom}-\cite{ACh}. In \cite{Gom} the CS
contribution was not taken into account for the case of massless fermions
(this in contrast with the results of
\cite{Jac3}, \cite{NS}-\cite{GMSS} where it was shown that even
massless fermions generate a CS term). In \cite{HP}, for an even
number of flavours, the total CS contribution was cancelled by
use of regularization sign ambiguities in a way that we find
rather artificial and that has been shown to be energetically unfavorable
in ref.\cite{ACh}.
Finally, our conclusions about parity violation and mass generation agree
with those obtained in \cite{ACh}, where the quartic interaction
was splitted using scalar fields and the resulting effective
theory was analysed within the $1/N$ expansion.

Let us end by noting that our approach can be straightforwardly
extended to the case of $U(N)$ Thirring model. We hope to report on
this issue in a separate work.

\underline{Acknowledgements}: The authors wish to thank A.\
Kovner for very helpful comments. G.R.\ is
partially supported by CONICET, Argentina. F.A.S.\ is partially
supported by CICBA, Argentina.

\end{document}